\documentstyle[pra,aps,eqsecnum,epsf]{revtex}
\def\ie{{\it i.e.,}\thinspace}
\newcommand{\beq}{\begin{equation}}
\newcommand{\eeq}{\end{equation}}
\newcommand{\beqa}{\begin{eqnarray}}
\newcommand{\eeqa}{\end{eqnarray}}
\def\displayfrac#1#2{\frac{\displaystyle #1}{\displaystyle #2}}
\begin{document}
\title{Bell's inequalities, multiphoton states and phase space distributions}
\author{Arvind\footnote[1]{email: arvind@physics.iisc.ernet.in}} 
\address{Department of Physics\\
Guru Nanak Dev University, Amritsar 142005, India}
\author{N. Mukunda\footnote[2]{ Also at Jawaharlal Nehru Centre for 
Advanced Scientific Research, Jakkur,
Bangalore - 560 064,India.}}
\address{Center for Theoretical Studies and Department of Physics\\
Indian Institute of Science,  Bangalore - 560 012, India}
\date{\today}
\maketitle 
\begin{abstract}
The connection between quantum optical nonclassicality and
the violation of Bell's inequalities is explored.  Bell
type inequalities for the electromagnetic field are formulated for 
general states of quantised radiation and their violation 
is connected to other nonclassical properties of the
field. This is achieved by considering states with an
arbitrary number of photons and carefully identifying the hermitian
operators whose expectation values do not admit any local
hidden variable description.  We relate the violation
of these multi-photon inequalities to  properties of phase space
distribution functions such as the diagonal coherent state
distribution function and the Wigner function. Finally, the
family of 4-mode states with Gaussian Wigner
distributions is analysed, bringing out in 
this case the connection  of violation of Bell type inequalities with 
the nonclassical property of squeezing.
\end{abstract}
\pacs{03.65.Bz,42.50.Dv}
\section{introduction}
The superposition principle and the reduction of the wave
function upon measurement are the two essential features of
quantum mechanics, which are responsible for most
counterintuitive paradoxical situations that arise when
one starts to interpret the predictions of quantum
mechanics. The violation of Bell's inequalities is 
one of quantum theory's most striking
consequences~\cite{bell-physics-64}.  When outcomes of
individual measurements are given an objective meaning, the
locality condition of ``no action at a distance'' imposes
constraints on the possible correlations, which are
expressed through Bell's inequalities.  There exist quantum
mechanical states for which these inequalities are violated
bringing out the fact that in quantum mechanics we either
have to give up hope of an objective interpretation of
individual measurements or accept ``action at a distance''.
The possibility of a local ``classical'' theory of
individual measurements is thus ruled out.  Though initial
work, starting with Bohm~\cite{bohm-pr-57} concerned itself with
the states of two spin-$\frac{1}{2}\/$ particles, it was
for particular states of the quantised electromagnetic field that
experiments were first performed in this
direction~\cite{clauser-prd-74}~\cite{aspect-prl-81}.

The formulation of Bell-type inequalities is possible only for
a system which has two or more kinematically independent subsystems.
These subsystems could each, for
example, be a spin-half system with states in a two dimensional 
Hilbert space, a photon with fixed energy-momentum but variable
polarisation state, or even a quantum mechanical system with one
canonically conjugate pair of operators $q\/$ and $p\/$
and states in an infinite dimensional Hilbert space~\cite{reid-pra-86}
\cite{rarity-prl-90}
\cite{ou-prl-92}.
Many discussions focus on states of two photons, for two fixed propagation 
vectors and variable polarization states. To go beyond a single photon 
in each mode, in our analysis we will deal fully with all states of 
4-mode fields, which include the two-photon states as a simple 
special case.
On the other hand, the two-mode electromagnetic 
fields, with different directions 
of propagation and fixed polarisations, provide us with a situation 
kinematically equivalent to the one 
considered in the EPR paper as the subsystems are of the 
$q\/, p\/$ form.

The quantum mechanical states for which Bell's inequalities
are violated thus have essential ``quantum'' features and
defy a deeper interpretation based on realism (the objective
existence of attributes independent of their measurement) and
locality (no action at a distance). On the other hand, for
a state which obeys a complete set of Bell's inequalities
it is in principle possible to give a local ``classical''
interpretation to individual measurements. An independent 
notion of classicality, used in quantum 
optics~\cite{walls-nature-79} \cite{klauder}, 
for the states of the quantised electromagnetic
 field is based on the diagonal 
coherent state distribution function, and here, the specific classical
theory one has in mind is Maxwell's theory. 
This paper explores the connections between these
two different ways of classifying the quantum mechanical
states of the electromagnetic field as ``nonclassical'' and
``classical''. To carry out such an analysis, we obviously have 
to go beyond two-photon states and formulate Bell type inequalities 
for general four mode states with arbitrary photon 
number distribution.

A typical setup used to study violation of Bell's
inequalities involves four modes of the field with
propagation in two different directions, and arbitrary
polarisations being allowed transverse to each direction.
For photons in each propagation direction  a 
particular polarisation is selected by a variable polariser,
and finally
coincidence counts are recorded using photo detectors. In
order to analyse an arbitrary state of the 4-mode field
and derive a Bell-type inequality for it, 
two crucial inputs are needed: the first is the
identification of hermitian operators whose measurements
and correlations are directly related to the coincidence
count rates; and the second is an appropriate quantum mechanical
analysis of the polariser, namely its action
on a general input multiphoton state.  We feel that previous analyses
in this direction lack proper treatment of one or the other
of these aspects~\cite{reid-pra-86}~\cite{anu-pra-91}~\cite{chubarov-pla-85}

A coincidence count may be said to be
registered when one or more photons are detected at each of
the two detectors simultaneously or within a pre-assigned
time interval, disregarding the exact numbers of
photons detected. (So this is a kind of coarse
grained coincidence rate). Hence such coincidence count rates 
differ from intensity-intensity
correlations. 
We introduce an appropriate set of hermitian
operators in the space of states of the 4-mode field
so that all such coincidence counts can be
calculated from the expectation values of products of these
operators. By assuming a  local realist hidden variable description for the 
outcome of various eigen values of the above mentioned operators
we derive Bell type inequalities constraining the
measurable coincidence count rates.

Classical states in the quantum optical sense have an
underlying classical distribution function; they have been
shown not to violate Bell's inequalities, while
nonclassical states are potential candidates for such
violations. The quantum optical nonclassicality is
invariant under passive canonical transformations which
form the group $U(n)\/$ for a general $n\/$ 
mode system~\cite{arvind-pramana-95} \cite{arvind-pra-95}; such
transformations can
be experimentally implemented in a straightforward way using
optically passive elements like beam splitters and mirrors.
On the other hand, such transformations can alter the potential
of a state to violate Bell type inequalities.
Therefore, starting with a nonclassical state we should
allow it to undergo arbitrary passive canonical
transformations before looking for violations of Bell's
inequalities. It may turn out that a state which obeys
Bell's inequalities is related by a passive transformation
to one which violates them; such a violation obviously
is a consequence of the nonclassicality of the original
state. This is closely connected to the fact that passive
transformations can take nonclassical nonentangled states
to nonclassical entangled ones.
Generally, therefore, the
relationship between total photon number conserving passive 
canonical transformations on the one hand, and entanglement or violation
of Bell-type inequalities on the other, has to be carefully analysed.
We show that this capacity of $U(n)\/$ to alter the entanglement properties
of states underlies the violation of Bell type inequalities by beams
originating from independent sources discussed by 
Yurke and Stoler~\cite{yurke-prl-92}~\cite{yurke-pra-92}.
Further we show that for their scheme to  work at least one
of the beams has to be in a quantum optically nonclassical state.
A much larger set of quantum mechanical states of the 4-mode electromagnetic 
field can be analysed in our formalism opening up the possibility 
of more easily experimentally
observing the predicted violations of Bell's inequalities. 

The material in this paper is arranged as follows: In
Section~II we describe the basic 4-mode setup to be used
for the study of Bell's inequalities. The coincidence count
rates for a general 4-mode state are then defined and used to
derive Bell type inequalities with an appropriate quantum
mechanical description for the polariser.  
Section~III
explores applications to various 4-mode states.
Coherent states which are ``classical'' in more than one
way (\ie they have classical diagonal coherent state
distribution functions, are not entangled in any basis and
lead to classically expected ``photon counting''), are shown to
obey Bell type inequalities.  This result is then used to 
prove that all quantum optically classical states also obey
our Bell-type inequalities. 
Finally we analyse in detail the family of states 
of the 4-mode field with a Gaussian-Wigner distribution, bringing
out their violation of the Bell-type inequalities and the connection of
such violation with the nonclassical property of squeezing.
Section~IV contains some concluding remarks.
\section{Bell's inequalities for multiphoton fields: choice
and preparation of states} 
Consider four modes of the electromagnetic field 
chosen so that we have two different propagation directions 
labeled by wave vectors ${\bf k}\/$ and ${\bf k}^{\prime}\/$
and along each direction we allow arbitrary
polarisations. We can choose for convenience a basis for
each polarisation space and can then label these modes with
annihilation operators
$a_{1},\/a_{2},\/a_{3}\/$ and
$a_{4}\/$;
the modes $a_{1}\/$ and $a_{2}\/$ refer to the linear
polarisations along $x\/$ and $y\/$ 
for the beam represented by wave vector ${\bf k}\/$ and the modes 
$a_{3}\/$ and  $a_{4}\/$ to  the linear polarisation modes  along
$x^{\prime}\/$ and $y^{\prime}\/$ for the 
wave vector ${\bf k}^{\prime}\/$.
Without any loss of generality, we can assume that both the
directions of propagation are in the plane of the paper;
then $x, x^{\prime}\/$ are also chosen to be 
in the same plane while $y,
y^{\prime}\/$ are in the common direction pointing out of the
plane of the paper. These are indicated in the left end of 
Figure~1. The $U(4)\/$ block symbolically represents the replacement
of $a_{1}, a_{2}, a_{3}, a_{4}\/$ by complex
linear combinations of themselves determined by a matrix of $U(4)\/$:
\beq
a_j^{\prime}= \sum_{k} U_{jk} a_k,\quad U \in U(4)
\eeq
which is a passive total photon number conserving canonical transformation.
Polarisers $P_1\/$ and $P_2\/$ can be
set at any angles $\theta_1\/$ and $\theta_2\/$ 
with respect to $x\/$ and $x^{\prime}\/$ respectively.
After passing through these polarisers the
beams encounter detectors $D_1\/$ and $D_2\/$ connected to
each other by a coincidence counter. 
This scheme is very
similar in spirit to most experimental and theoretical
situations which have been extensively investigated.   
\begin{figure}[h]
\hspace*{1.5cm}\epsfxsize=14cm 
\epsfbox{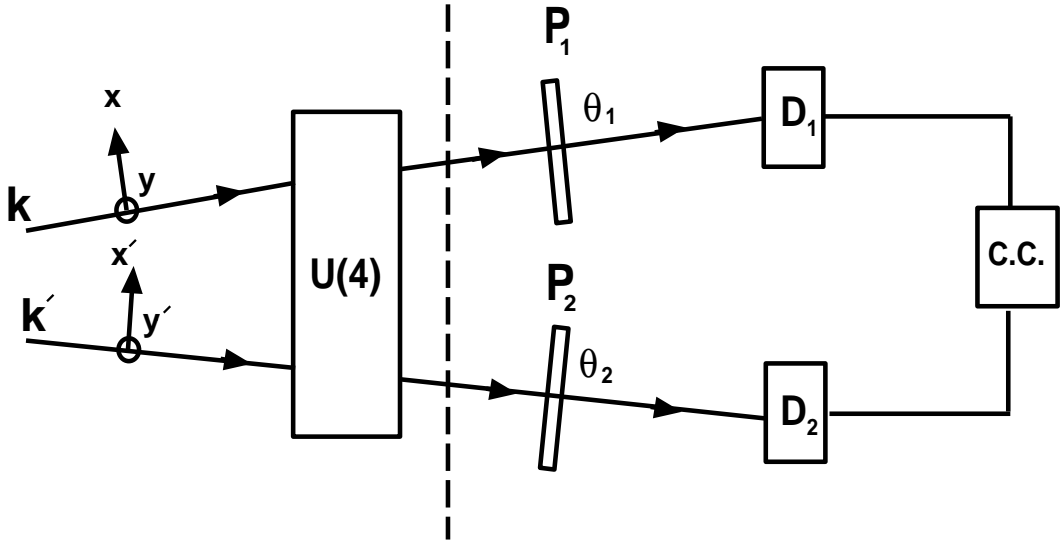}
\vspace*{1cm}
\caption{Setup to study the violation of Bell
type inequalities for arbitrary states of the four mode radiation
field. The {\bf U(4)} block represents a variable passive
$U(4)\/$ canonical transformation mixing the four modes, 
$P_1, P_2\/$ are the polarisers oriented at
angles $\theta_1, \theta_2\/$ with respect to the axes
$x\/$ and $x^{\prime}\/$ respectively. D${}_1\/$ and
D$_{2}\/$ represent the detectors and the block c.c.
indicates the coincidence counter. The part of the diagram to the left of 
the vertical line indicates symbolically the state preparation stage,
that to the right the measurements done on it.}
\end{figure}

However, the difference here is that we do not wish to
assume anything about the incident 4-mode state; it
could have {\em arbitrary number of photons},
and could even be a {\em mixed state}. The
measurements of interest here are the coincidence count
rates, either for given settings of the polarisers (\ie
$\theta_1\/$ and $\theta_2\/$) or when one or both
polarisers are removed. For a given $\theta_1(\theta_2)\/$,
$D_1(D_2)\/$ receives and detects photons in the
corresponding linear polarisation state alone.(For
simplicity alone we restrict the analysis here to linear
polarisation states) 
\subsection{Interpretation of the coincidence count rates}
A coincidence is counted when both the detectors $D_1\/$
and $D_2\/$ click simultaneously(`Simultaneous' here has 
the meaning as stated in the
introduction that within a
preassigned time interval $\Delta t\/$ each of the two
detectors registers at least one photon).
We will not distinguish
in this analysis between coincidences of different
strengths \ie different numbers of photons (greater than
one) being received by each
detector. Coincidence rates defined in this specific physical manner
must be represented by suitable choices of hermitian operators and their
expectation values; this will be done in detail below.
In our analysis, we will need the following four
types of coincidence count rates:
\begin{itemize}
\item[(a)]
$P(\theta_1,\theta_2)\/$: The first polariser at 
$\theta_1\/$ and the second one at $\theta_2\/$ with
respect to their respective $x\/$ axes.
\item[(b)]
$P(\theta_1,\quad)\/$: The first polariser at
$\theta_1\/$ and the second one  removed. 
\item[(c)]
$P(\quad,\theta_2)\/$: The first polariser removed and
the second one at $\theta_2\/$.
\item[(d)]
 $P(\quad,\quad)\/$: Both the polarisers removed from the setup.
\end{itemize}
These coincidence counts are the measurable quantities and
can be calculated for any quantum mechanical state of the
4-mode field. On the other hand, we will see that these
count rates have to obey certain inequalities if we demand
a realist description and locality as well, so 
that measurements at detector $D_1\/$ are independent of the
setting of the polariser $P_2\/$ and vice
versa. 

\subsection{Quantum mechanical description of the polariser}
Classically, the action of a polariser is straightforward.
It allows a particular polarisation state (which defines
the polariser axis) to pass through without any hindrance
and blocks the orthogonal one completely. For a general
polarisation state, the component of the electric field
along the axis passes through unaffected while the
orthogonal component is absorbed. We are interested in the
generalisation of this action to quantum mechanical
situations. For a comparable discussion for the beam splitter
action see~\cite{campos-pra-89}.
Given the classical action, we know how the
polariser should act on single photons and on coherent
states. For a single photon traveling in the $z\/$
direction, in an arbitrary state of polarisation
$\/c_1\vert 1 \rangle_x
\vert 0 \rangle_y +\/ c_2\vert 0 \rangle_x \vert 1 \rangle_y 
\/$ ($c_1, c_2\/$ being
arbitrary complex constants with $\vert c_1 \vert^2 + \vert
c_2 \vert^2 =1$)\footnote[2]{Here we have used the
language appropriate to two modes 
and allowed the single photon to be in either
of the polarisation modes; equivalently we could have
talked about a single photon in a superposition of two
orthogonal polarisation states} in the $x-y\/$ plane and
with a polariser placed in the $x\/$ direction, the
probability of transmission is $\vert c_1\vert^2\/$. From
this we can build up the probabilities of transmission for
arbitrary states as they can be expanded in terms of number
states. Consider a general pure input state 
$\sum_{n_1,n_2} C_{n_1n_2} \vert n_1 \rangle_{x}
\vert n_2 \rangle_{y}\/$ with the polariser placed along
the $x\/$ direction. The probability of finding $n_1\/$ photons
after passage through the polariser should be given by
$ \sum_{n_2} \vert C_{n_1n_2}\vert^2\/$. 
Nevertheless, this still leaves the question about
the output state open-ended; what is the state of the single-mode
field which emerges out of the polariser?  Exactly
identical probabilities can be obtained from pure as well
as mixed states emerging from the polariser. Compared to
the classical case, we need to give more detailed
consideration to the physical process underlying the
removal of photons of particular polarisation from the beam; it
could either be absorption by certain other degrees of
freedom(lattice) or a change of direction caused by
different refractive indices for different polarisations.
In either case, after passage through the polariser the
information contained in that mode, though existing, is
inaccessible and therefore we should trace over that mode
to obtain the outgoing state of the field.

We thus arrive at the following general prescription for
the action of the polariser: for a given input two-mode state
with density matrix $\rho\/$, the two
modes being two orthogonal polarisations along the same
direction of propagation, and a polariser placed at an angle
$\theta \/$ with respect to the $x\/$ axis, the single-mode
state $\rho(\theta)\/$ after passage through the polariser
is obtained by taking the trace over the mode orthogonal to
the linear polarisation defined by $\theta\/$:
\beq
\rho(\theta) = 
\sum \limits_{n=0}^\infty {}_{(\theta+\frac{\pi}{2})}\langle n\vert \rho
\vert n \rangle_{(\theta+\frac{\pi}{2})}
\eeq
Here we have chosen the number state basis for the
mode orthogonal to $\theta\/$; we could as well have chosen
any other complete set of states.
More explicitly, consider the most general density matrix for
the two polarisation modes along the same direction of propagation. 
\beq
\rho = \sum_{n_1,n_2,n^{\prime}_1,n^{\prime}_2} 
C_{n_1 n_2 n^{\prime}_1 n^{\prime}_2} \vert n_1 \rangle_x
\vert n_2 \rangle_y \,{}_x\langle n_1^{\prime} \vert
\,{}_y\langle n_2^{\prime} \vert 
\eeq 
If a polariser is placed at an angle $\theta\/$ with
respect to the $x\/$ axis we first rotate the basis in the
$x-y \/$ plane to $\theta\/$ and $\theta +\frac{\pi}{2}\/$ 
giving us
\beq
\rho = \sum_{n_1,n_2,n^{\prime}_1,n^{\prime}_2} 
C_{n_1 n_2 n^{\prime}_1 n^{\prime}_2}(\theta) \vert n_1 \rangle_{\theta}
\vert n_2 \rangle_{\theta + \frac{\pi}{2}} \,{}_{\theta}
\langle n_1^{\prime} \vert
\,{}_{\theta+\frac{\pi}{2}}\langle n_2^{\prime} \vert 
\eeq
Each $C_{n_1 n_2 n^{\prime}_1 n^{\prime}_2}(\theta)\/$
is a $\theta \/$ dependent linear expression in the
$C_{....}$'s, of course
conserving $n_1+n_2$ and $n_1^{\prime}  + n_2^{\prime}\/$.
Now tracing over the mode orthogonal to $\theta\/$ yields
the final state after the passage through the polariser: 
\beq
\rho(\theta) = \sum_{n_1,n_1^{\prime}}
\left(\sum_{n_2}C_{n_1 n_2 n^{\prime}_1
n_2}(\theta) \right) \vert n_1 \rangle_{\theta}
\,{}_{\theta}\langle n_1^{\prime} \vert
\eeq
When $\rho\/$ is a pure 
state  \ie 
$C_{n_1 n_2 n^{\prime}_1 n^{\prime}_2}=C_{n_1 n_2}
C^{\star}_{n^{\prime}_1 n^{\prime}_2},\/$ the rotated
coefficients  
$C_{n_1 n_2 n^{\prime}_1 n^{\prime}_2}(\theta)\/$ also
factorise, and the output state 
\beq
\rho(\theta) = \sum_{n_1,n_1^{\prime}}
\left(\sum_{n_2}C_{n_1 n_2}(\theta) C^{\star}_{n^{\prime}_1 n_2}(\theta)
\right) \vert n_1 \rangle_{\theta}
\,{}_{\theta}\langle n_1^{\prime} \vert
\eeq
could in general be mixed. For the special case of coherent
states $\vert z_1\rangle_x \vert z_2 \rangle_y\/$
\beqa
C_{n_1 n_2}&=&e^{\displaystyle -\frac{1}{2}(\vert z_1\vert^2 + \vert
z_2\vert^2)}\; \displayfrac{z_1^{n_1}}{\sqrt{n_1!}}
\displayfrac{z_2^{n_2}}{\sqrt{n_2!}}
\nonumber \\
C_{n_1 n_2}(\theta)&=&
e^{-\displaystyle \frac{1}{2}(\vert z_1\vert^2 + \vert
z_2\vert^2)} \;
\displayfrac{(z_1 \cos{\theta}-z_2 \sin{\theta})^{n_1}}{\sqrt{n_1!}} 
\displayfrac{ (z_1 \sin{\theta}+z_2 \cos{\theta})^{n_2}}{\sqrt{n_2!}}
\eeqa
Therefore the final density matrix after the beam emerges
from the polariser is given by
\beqa
&&\!\!\!\!\!\!\!\! \rho_{cs}(\theta) =
e^{\displaystyle -(\vert z_1\vert^2 + \vert
z_2\vert^2)} \;
\sum_{n_1,n_1^{\prime}}
\displayfrac{(z_1 \cos{\theta}-z_2 
\sin{\theta})^{n_1}}{\sqrt{n_1!}} \times
\nonumber \\
&&\quad \quad \quad\quad\quad \quad
\displayfrac{(z^{\star}_1 \cos{\theta}-z^{\star}_2 
\sin{\theta})^{n^{\prime}_1}}{\sqrt{n^{\prime}_1!}}
\,
\sum_{n_2}\displayfrac{ 
(\vert z_1 \sin{\theta}+z_2 \cos{\theta}\vert)^{2
n_2}}{n_2!} 
\vert n_1 \rangle_{\theta}\,{}_\theta\langle n^{\prime}_1 \vert
\nonumber \\
&&\!\!\!\!= 
e^{\displaystyle -\vert z_1 \cos{\theta}-z_2 \sin{\theta}\vert^2}\;
\sum_{n_1,n_1^{\prime}}
\displayfrac{(z_1 \cos{\theta}-z_2 \sin{\theta})^{n_1}}{\sqrt{n_1!}}
\,
\displayfrac{(z^{\star}_1 \cos{\theta}-z^{\star}_2 
\sin{\theta})^{n^{\prime}_1}}{\sqrt{n^{\prime}_1!}}
\vert n_1 \rangle_{\theta}\,{}_\theta\langle n^{\prime}_1\vert
\eeqa
which is a density matrix corresponding to the pure coherent
state $\vert z_1 \cos{\theta}-z_2 \sin{\theta} \rangle_{\theta} \/$,
consisting of photons all of which are polarized in the
direction $\theta\/$.
The reason for
this is the subtle fact that coherent states are not
entangled in any basis we may choose for the two-mode
incident state, and thus tracing over one of the modes just
amounts to neglecting the corresponding factor in the
product state after an appropriate rotation of the basis
\beq
\vert z_1 \rangle_x \vert z_2 \rangle_y =
\vert z_1\cos{\theta}-z_2 \sin{\theta} \rangle_{\theta}
\vert z_1\sin{\theta}+z_2 \cos{\theta}
\rangle_{\theta+\frac{\pi}{2}} 
\longrightarrow \vert z_1 \cos{\theta}-z_2 \cos{\theta}
\rangle_{\theta}  
\eeq
On the other hand, single photon states can in general be
entangled states of the two-mode field 
and thus would lead to mixed one-mode
states after passing through the polariser. For
example a pure two-mode single photon state
$\frac{1}{\sqrt{2}}(\vert 1
\rangle_x \vert 0\rangle_y+\vert 0
\rangle_x \vert 1 \rangle_y)\/$, after passage through a
polariser placed in the $x\/$ direction reduces 
to a mixed state with density matrix 
$\frac{1}{2}(\vert 0 \rangle_x \, {}_x \langle 0\vert + 
\vert 1 \rangle_x \,{}_x \langle 1\vert)$. 
\subsection{Derivation of the inequalities}
In order to define and  calculate the coincidence count rates, 
consider the following four hermitian operators, all having
eigen values $0\/$ and $1\/$
\beqa
\widehat{A}_1 = \left( I_{2\times2} - \vert 0 0 \rangle
\langle 0 0 \vert \right)_{\bf k} \nonumber \\
\widehat{A}_2 = \left( I_{2\times2} - \vert 0 0 \rangle
\langle 0 0 \vert \right)_{{\bf k}^{\prime}} \nonumber \\
\widehat{A}_1(\theta_1) = \left( I_{\theta_1} - 
\vert 0  \rangle_{\!\theta_1}
\,{}_{\theta_1\!}\langle 0 \vert \right)I_{\theta_1+\frac{\pi}{2}} \nonumber \\
\widehat{A}_2(\theta_2) = \left( I_{\theta_2} - 
\vert 0  \rangle_{\!\theta_2}
\,{}_{\theta_2\!}\langle 0 \vert \right)I_{\theta_2+\frac{\pi}{2}} 
\label{A-operators}
\eeqa
The subscripts $\theta_1\/$ and $\theta_2\/$ in the last
two equations refer to the directions of the polarisers. 
Thus $\widehat{A}_1\/$ and $\widehat{A}_1(\theta_1)\/$ 
are operators belonging to
the first two modes of our 4-mode system, namely
propagation direction ${\bf k}\/$ and polarisation along $x\/$ or
$y\/$; $I_{2\times2}\/$ is the unit operator for these two
modes while $I_{\theta_1}\/$ is the unit operator for the
single mode propagating in direction ${\bf k}\/$ and with
polarisation $\theta_1\/$, and $I_{\theta_2 +
\frac{\pi}{2}}\/$ that for the orthogonal polarisation.
$\widehat{A}_2\/$ and $\widehat{A}_2(\theta_2)\/$ 
are defined in a similar way
for propagation direction ${\bf k}^{\prime}\/$. 
Expectation values of the
above operators are probabilities of finding at least one
photon of the appropriate kind:
\beqa
\langle \widehat{A}_1 \rangle &=&
\mbox{probability of detecting at least one photon at
$D_1\/$ with $P_1\/$ removed,}
\nonumber \\
\langle \widehat{A}_2 \rangle &=& 
\mbox{probability of detecting at least one photon at
$D_2\/$ with $P_2\/$ removed,}
\nonumber \\
\langle \widehat{A}_1(\theta_1) \rangle &=&
\mbox{probability of detecting at least one photon at
$D_1\/$ with $P_1\/$ set at $\theta_1\/$,}
\nonumber \\
\langle \widehat{A}_2(\theta_2) \rangle &=& 
\mbox{probability of detecting at least one photon at
$D_2\/$ with $P_2\/$ set at $\theta_2\/$.}
\eeqa
These operators are 
defined so that when we calculate their expectation values in any
state, 
the actions of the polarisers as described in
the previous subsection are automatically implemented.
In an individual measurement, one of the
eigen values of the hermitian operator is recorded, with the
probabilities for the two possible eigen values being
calculable from the wavefunction or the density matrix. 
On the other hand if a
hidden variable description is available, then given the
value of the hidden variable $\lambda\/$ as well as the
quantum mechanical state vector, we can in principle
predict the outcomes of individual measurements. It is well
known that while keeping the hidden variable theory
completely general and by imposing the locality condition
(no action at a distance), general constraints on the
correlations between  observed quantities can be
obtained; we will perform a similar analysis for our
situation and derive Bell type inequalities for the
correlations among the $\widehat{A}\/$'s.

Assuming the probability distribution for $\lambda\/$ to be
${\cal P}(\lambda)\/$(for simplicity the dependence of 
${\cal P}(\lambda)\/$ on the specific 4-mode quantum state is omitted)
so that 
\beq
\int {\cal P}(\lambda)\/ d \lambda =1,
\eeq
we now proceed to develop expressions for
the average values of various
observables. Let $a_1(\lambda), a_2(\lambda), a_1(\theta_1,
\lambda),$ and $a_2(\theta_2, \lambda)\/$ be the actual
values of the dynamical variables $A_1, A_2,
A_1(\theta_1)\/$ and $A_2(\theta_2)\/$ for a given value of
$\lambda \/$ respectively: they take values $0\/$ or
$1\/$. The averages computed from hidden variable
theory are then given by
\beqa
\langle A_1 \rangle_{\rm hv} &=& \int a_1(\lambda)\/
{\cal P}(\lambda)\/ d \lambda \nonumber \\
\langle A_2 \rangle_{\rm hv} &=& \int a_2(\lambda)\/
{\cal P}(\lambda)\/ d \lambda \nonumber \\
\langle A_1(\theta_1) 
\rangle_{\rm hv} &=& \int a_1(\theta_1,\lambda)\/
{\cal P}(\lambda)\/ d \lambda \nonumber \\
\langle A_2(\theta_2)
\rangle_{\rm hv} &=& \int a_2(\theta_2,\lambda)\/
{\cal P}(\lambda)\/ d \lambda
\eeqa
The average values of the products of these operators, 
corresponding
 to the various coincidence count rates, are then given in
local hidden variable theory by
\beqa
P(\quad,\quad)_{\rm hv}=
\langle A_1 A_2 \rangle_{\rm hv} &=& \int a_{12}(\lambda)\/
{\cal P}(\lambda)\/ d \lambda =\int a_1(\lambda) a_2(\lambda)\/
{\cal P}(\lambda)\/ d \lambda  \nonumber \\
P(\theta_1,\quad)_{\rm hv}=
\langle A_1(\theta_1) A_2 \rangle_{\rm hv} &=&
\int a_{12}(\theta_1, \lambda)\/
{\cal P}(\lambda)\/ d \lambda =\int a_1(\theta_1,\lambda) a_2(\lambda)\/
{\cal P}(\lambda)\/ d \lambda  \nonumber \\
P(\quad,\theta_2)_{\rm hv}=
\langle A_1 A_2(\theta_2)
 \rangle_{\rm hv} &=& \int a_{21}(\theta_2,\lambda)\/
{\cal P}(\lambda)\/ d \lambda =\int a_1(\lambda) a_2(\theta_2,\lambda)\/
{\cal P}(\lambda)\/ d \lambda  \nonumber \\
P(\theta_1,\theta_2)_{\rm hv}=
\langle A_1(\theta_1) A_2(\theta_2)
 \rangle_{\rm hv} &=& \int a_{12}(\theta_1,\theta_2,\lambda)\/
{\cal P}(\lambda)\/ d \lambda =\int a_1(\theta_1,\lambda) 
a_2(\theta_2,\lambda)\/
{\cal P}(\lambda)\/ d \lambda \nonumber \\
\label{product-form}
\eeqa
The subscript ``hv'' is for `hidden variable'
 and
distinguishes these averages from the quantum mechanical
expectation values.  Here in the first step, $a_{12}$'s are
the hidden variable theory values of corresponding dynamical
variables $A_1 A_2\/$ etc.  The last step in each of the
above equations is very crucial and is basically the
expression of the locality assumption on the hidden
variable theory. We have assumed for instance 
that the quantity $a_{12}(\theta_1,\theta_2,\lambda)\/$ 
is a product of the two factors
$a_1(\theta_1, \lambda)\/$ and $a_2(\theta_2, \lambda)\/$,
each depending only on one angle and not the other; more
generally what we measure for the propagation direction ${\bf k}\/$
does not depend on what we choose to measure (or not
measure) for direction ${\bf k}^{\prime}\/$.
This assumption is ``reasonable''
because we can arrange the situation such that the two
measurement events are space like separated.

One more assumption we  make is the ``no
enhancement assumption''\footnote[1]
{This is at the level of hidden variable description. We
have already seen that the possible values of
$a_1(\theta_1, \lambda)\/$ and $a_1\/$ are $0\/$ or $1$. So if
$a_1(\lambda) = 0\/$ for some $\lambda\/$ then
$a_1(\theta_1, \lambda) =0 \/$  as well for any $\theta_1\/$. }
\ie the presence of the polariser
can only remove photons from the beam and is incapable of
adding photons or increasing the
coincidence count rate:
\beqa
a_1(\theta_1,\lambda)  \leq a_1(\lambda) \nonumber\\
a_2(\theta_2,\lambda)  \leq a_2(\lambda)
\label{noenhance}
\eeqa
We now recall the lemma due to Clauser and
Horne~\cite{clauser-prd-74}  which we
will use to derive inequalities for the above calculated
correlation functions.\\
\underline{\bf lemma:}\\
if $ 0\leq \/x, x^{\prime}\/ \leq X\/$ and $ 0\leq \/ y,
y^{\prime} \leq Y\/$ then
\beq
-XY \leq x y - x y{\prime} + x^{\prime} y + x^{\prime}
y^{\prime} - Y x^{\prime} - X y \leq 0
\label{lemma}
\eeq

With the ``no enhancement assumption''~(\ref{noenhance}) and the
product forms~(\ref{product-form}), after identifying
$x=a_1(\theta_1, \lambda),\,x^{\prime}=a_1(\theta^{\prime}_1,\, \lambda),
y=a_2(\theta_2, \lambda),\, y^{\prime}=a_2(\theta^{\prime}_2, \lambda),\,
X=a_1(\lambda),\,  Y=a_2(\lambda)\/$ we get the following
inequality
\beqa
-a_1(\lambda) a_2(\lambda)\/ &\leq\/&
a_1(\theta_1, \lambda) a_2(\theta_2,\lambda)
-a_1(\theta_1, \lambda) a_2(\theta_2^{\prime},\lambda)
+a_1(\theta_1^{\prime}, \lambda) a_2(\theta_2,\lambda)
+a_1(\theta_1^{\prime}, \lambda) a_2(\theta_2^{\prime},\lambda)
\nonumber \\
&&\quad\quad\quad- a_2(\lambda) a_1(\theta_1^{\prime},\lambda)
-a_1(\lambda) a_2(\theta_2,\lambda)\/ \leq\/ 0
\eeqa
Integrating the above inequality over $\lambda\/$, with
weight function ${\cal P}(\lambda)\/$ we get the inequality
obeyed by the coincidence count rates for any choices of 
angles $\theta_1, \theta_2, \theta^{\prime}_1, \theta^{\prime}_2\/$
\beq
-P(\quad,\quad)\/ \leq\/
P(\theta_1,\theta_2)
-P(\theta_1,\theta_2^{\prime})
+P(\theta_1^{\prime},\theta_2)
+P(\theta_1^{\prime},\theta_2^{\prime})
-P(\theta^{\prime}_1,\quad)
-P(\quad,\theta_2)
 \/\leq\/ 0
\label{chs-ineq}
\eeq 
This is the inequality we will use in our future analysis.
The left hand side inequality depends upon the total
coincidence count rate without the polarisers whereas the
one on the right hand side does not. This result is of the same
form as that derived and used by all workers on the subject;
what we emphasize here is that we have clearly specified the 
hermitian operator observables at the quantum mechanical level, for
which correlations are considered, so the actual expressions for
various $P$'s are specific to our treatment.

For such a theory to be consistent with quantum mechanics,
the coincidence rates computed from a quantum mechanical
calculation should also obey the above inequalities.
In the above derivation we have not assumed anything about
the 4-mode state and therefore we can consider
arbitrary states to check for any violation. 

If a given quantum mechanical state does not obey the above
inequalities then it definitely has some nontrivial quantum
features which cannot be accommodated within realist hidden variable
theories based on locality. We now discuss various examples of 
4-mode quantum states in sequence, and examine the validity or
otherwise of inequalities~(\ref{chs-ineq}) in each case.
\section{Multiphoton states, violation of Bell's inequalities and
phase space distributions}
\subsection{Two-photon states}
Before undertaking the analysis of multiphoton states, 
we consider here the extensively studied two photon state
as a warm up exercise. It so happens that for 
these states our formalism  
reduces to the usual one and we get results
identical to those already available in the literature. 
Consider the following pure two-photon state 
of the 4-mode field
\beqa
\vert \psi \rangle &=& 
\frac{1}{2}(
\vert 1 \rangle_{1}\vert 0 \rangle_{2}
\vert 0 \rangle_{3}\vert 1 \rangle_{4} -
\vert 1 \rangle_{1}\vert 1 \rangle_{2}
\vert 0 \rangle_{3}\vert 0 \rangle_{4} -
\vert 0 \rangle_{1}\vert 0 \rangle_{2}
\vert 1 \rangle_{3}\vert 1 \rangle_{4} +
\vert 0 \rangle_{1}\vert 1 \rangle_{2}
\vert 1 \rangle_{3}\vert 0 \rangle_{4}) \nonumber \\
&=& \frac{1}{2}
(a^{\dagger}_{1}-a^{\dagger}_{3})
(a^{\dagger}_{4}-a^{\dagger}_{2}) 
\vert 0 \rangle_{1}\vert 0 \rangle_{2}
\vert 0 \rangle_{3}\vert 0 \rangle_{4}
\eeqa

The calculations of the quantum mechanical 
coincidence count rates become easy
if we make the following observation about the operators
$\widehat{A}_1(\theta_1)\/$ and $\widehat{A}_2(\theta_2)\/$:
\beqa
\widehat{A}_1(\theta_1)\/= 
{\cal U}(R_1(\theta_1))
\,(I_{1} - \vert 0 \rangle_{1} \,{}_{1}\langle 0 \vert)
\,I_{2}\,{\cal U}^{-1}(R_1(\theta_1))
\nonumber \\
\widehat{A}_2(\theta_2)\/= 
{\cal U}(R_2(\theta_2))\,
(I_{3} - 
\vert 0 \rangle_{3} \,{}_{3}\langle 0 \vert)
\,I_{4}\,{\cal U}^{-1}(R_2(\theta_2))
\eeqa
Here ${\cal U}(R_1(\theta_1))\/$ and ${\cal U}(R_2(\theta_2))\/$
are the unitary operators corresponding to the
transformations $R_1(\theta_1)\/$ and $R_2(\theta_2)\/$ which
rotate the basis in the polarisation space by
angles $\theta_1\/$ and $\theta_2\/$ for directions ${\bf k}\/$ 
and ${\bf k}^{\prime}\/$
respectively.
\beq
R_1(\theta_1)=\left(\begin{array}{cc}
\cos{\theta_1} & \sin{\theta_1}\\
-\sin{\theta_1} & \cos{\theta_1}
\end{array} \right), \quad 
R_2(\theta_2)=\left(\begin{array}{cc}
\cos{\theta_2} & \sin{\theta_2}\\
-\sin{\theta_2} & \cos{\theta_2}
\end{array} \right)
\label{pol}
\eeq
Using this form of the operators the calculation of the
expectation values of various combinations of operators is
straight forward and yields the quantum mechanical
 coincidence count rates:
\beqa
P(\theta_1,\theta_2)_{\mbox{qm}}^{\mbox{tp}} &=& 
\langle \psi \vert \widehat{A}_1(\theta_1)
\widehat{A}_2(\theta_2) \vert \psi \rangle
=\displayfrac{1}{4} \sin^2{(\theta_1+\theta_2)}\nonumber \\
P(\theta_1, \quad)_{\mbox{qm}}^{\mbox{tp}}&=& \langle \psi
\vert \widehat{A}_1(\theta_1)
\widehat{A}_2 \vert \psi \rangle =\displayfrac{1}{4} \nonumber \\
P(\quad,\theta_2)_{\mbox{qm}}^{\mbox{tp}}&=& \langle \psi \vert \widehat{A}_1
\widehat{A}_2(\theta_2) \vert \psi \rangle =\displayfrac{1}{4}\nonumber\\
P(\quad, \quad)_{\mbox{qm}}^{\mbox{tp}} &=&\langle \psi \vert \widehat{A}_1
\widehat{A}_2 \vert \psi \rangle= \displayfrac{1}{2}
\eeqa
(The superscript `tp' here refers to two-photon states).
Substituting these rates in the
inequality~(\ref{chs-ineq}) we get the following condition,
if this state is capable of a local realist hidden variable
description:
\beq
-1 \/\leq 
\displayfrac{1}{2}(\sin^2(\theta_1+\theta_2)
-\sin^2(\theta_1+\theta^{\prime}_2)
+\sin^2(\theta^{\prime}_1+\theta_2)
+\sin^2(\theta^{\prime}_1+\theta^{\prime}_2)
-2)\/ \leq 0
\eeq
We find that 
the above condition can be violated on either side for some 
values of $\theta_1\/,\theta_1^{\prime},\theta_2$ and 
$\theta_2^{\prime}\/$. As an example
let us choose $\theta_1=\displayfrac{\pi}{8},
\theta_2=\displayfrac{\pi}{4},
\theta^{\prime}_1=\displayfrac{3 \pi}{8}\/$ and
 $ \theta_2^{\prime}=0\/$; 
 then the right hand side of the above inequality becomes 
\beq
\displayfrac{1}{2}\/(\sqrt{2}-1) \leq 0
\eeq
which is clearly violated.

Since two-photon states have been extensively studied in the
literature we connect our result with existing analyses and 
make some useful observations. The single hermitian operator
$\widehat{A}=I-\vert 0 \rangle \langle 0 \vert \/$ whose expectation
value gives the probability for finding one or more photons
reduces effectively 
to $a^{\dagger}a\/$ for the appropriate mode when
at most one photon is present in the beam; therefore our
results agree with existing ones~\cite{ou-prl-88}. 
For two-photon states we have the relations
$P(\theta_1, \quad) = P(\theta_1, \theta_2)+ P(\theta_1,
\theta_2+ \frac{\pi}{2})\/$ etc. They too are a consequence
of the reduction of the $\widehat{A}\/$ 
operators to the above described form
and do not remain valid for the analysis of general
states.

\subsection{The coherent states}
As a first nontrivial example we consider 
4-mode coherent states defined as (omitting further
subscripts outside kets)
\beqa
\vert z_{1} \rangle \vert z_{2} \rangle
\vert z_{3} \rangle \vert z_{4} \rangle &&=
{\displaystyle e}^{-\displayfrac{1}{2}\left(
\displaystyle \vert z_{1} \vert^2
    +\vert z_{2} \vert^2
    +\vert z_{3} \vert^2
    +\vert z_{4} \vert^2\right)} \times
\nonumber \\
&&{\displaystyle e}^{\displaystyle  
\left(z_{1} a^{\dagger}_{1}
         + z_{2} a^{\dagger}_{2}
         + z_{3} a^{\dagger}_{3}
         + z_{4} a^{\dagger}_{4} \right)}
\vert 0 0 0 0 \rangle
\eeqa
where $z_{1}, z_{2}, z_{3}, z_{4}\/$ are complex numbers.
The calculation of quantum mechanical 
coincidence count rates for this case is
rather straightforward and we get the following results
(the superscript `cs' means coherent states):
\beqa
P(\theta_1,\theta_2)_{\mbox{qm}}^{\mbox{cs}} &=& (1-e^{\displaystyle 
-\vert z^{\prime}_{1}\vert^2})
                          (1-e^{\displaystyle 
-\vert z^{\prime}_{3}\vert^2})
\nonumber \\
P(\theta_1,\quad)_{\mbox{qm}}^{\mbox{cs}} &=& 
(1-e^{\displaystyle -\vert z^{\prime}_{1}\vert^2})
(1-e^{\displaystyle  -(\vert z_{3}\vert^2 +
\vert z_{4}\vert^2)})
\nonumber \\
P(\quad,\theta_2)_{\mbox{qm}}^{\mbox{cs}} &=& 
(1-e^{\displaystyle -\vert z^{\prime}_{3}\vert^2})
(1-e^{\displaystyle -(\vert z_{1}\vert^2 +
\vert z_{2}\vert^2)})
\nonumber \\
P(\quad,\quad)_{\mbox{qm}}^{\mbox{cs}} &=&
(1-e^{\displaystyle 
-(\vert z_{1}\vert^2 +\vert z_{2}\vert^2)})
(1-e^{\displaystyle 
-(\vert z_{3}\vert^2 +\vert z_{4}\vert^2)})
\nonumber  \\
z_{1}^{\prime} &=& z_{1} \cos{\theta_1} - z_{2}
\sin{\theta_1} 
\nonumber \\
z_{2}^{\prime} &=& z_{1} \sin{\theta_1} + z_{2}
\cos{\theta_1}
 \nonumber \\
z_{3}^{\prime} &=& z_{3} \cos{\theta_2}
- z_{4} \sin{\theta_2}
\nonumber  \\
z_{4}^{\prime} &=& z_{3} 
\sin{\theta_2} + z_{4} \cos{\theta_2}
\label{coh-coincidence}
\eeqa
We see that $P^{\rm cs}_{\rm qm}(\theta_1, \theta_2)\/$ is a product of two
factors, the first one depending on $\theta_1\/$ alone and the second
on $\theta_2\/$ alone. This fact originates
from the unentangled nature of the coherent states and will
be used now to show that coherent states always obey the
inequalities~(\ref{chs-ineq}). Using the facts that 
$0 \/\leq\/ P(\theta_1, \quad)_{\mbox{qm}}^{\mbox{cs}} , 
P(\theta_1^{\prime},\quad)_{\mbox{qm}}^{\mbox{cs}},
P(\quad,\theta_2)_{\mbox{qm}}^{\mbox{cs}}, 
P(\quad,\theta_2^{\prime})_{\mbox{qm}}^{\mbox{cs}}
\/\leq\/ P(\quad, \quad)_{\mbox{qm}}^{\mbox{cs}} \/$ 
and the lemma~(\ref{lemma}),
with the identification 
$x=P(\theta_1,\quad)_{\mbox{qm}}^{\mbox{cs}} ,\,
x^{\prime}=P(\theta^{\prime}_1,\quad)_{\mbox{qm}}^{\mbox{cs}} ,\,
y=P(\quad,\theta_2)_{\mbox{qm}}^{\mbox{cs}} ,\,
y^{\prime}=P(\quad,\theta^{\prime}_2)_{\mbox{qm}}^{\mbox{cs}} ,\,
X=Y=P(\quad,\quad)_{\mbox{qm}}^{\mbox{cs}} \/$
we arrive at the following preliminary inequality
\beqa
&&-P(\quad,\quad)^{2}_{\mbox{qm}}{}^{\mbox{cs}}\/ \leq \/ \nonumber \\
&&\quad\quad P(\theta_1,\quad)_{\mbox{qm}}^{\mbox{cs}}
P(\quad,\theta_2)_{\mbox{qm}}^{\mbox{cs}}
-P(\theta_1,\quad)_{\mbox{qm}}^{\mbox{cs}} 
P(\quad,\theta^{\prime}_2)_{\mbox{qm}}^{\mbox{cs}}
+P(\theta^{\prime}_1,\quad)_{\mbox{qm}}^{\mbox{cs}}
P(\quad,\theta_2)_{\mbox{qm}}^{\mbox{cs}} \nonumber \\
&&\quad\quad+P(\theta^{\prime}_1,\quad)_{\mbox{qm}}^{\mbox{cs}} 
P(\quad,\theta^{\prime}_2)_{\mbox{qm}}^{\mbox{cs}}
-P(\theta_1^{\prime},\quad)_{\mbox{qm}}^{\mbox{cs}} 
P(\quad,\quad)_{\mbox{qm}}^{\mbox{cs}}
-P(\quad,\theta_2)_{\mbox{qm}}^{\mbox{cs}}
P(\quad,\quad)_{\mbox{qm}}^{\mbox{cs}} 
\nonumber \\
 &&\quad\quad\quad\quad\/ \leq \/ 0
\label{coherent-lemma}
\eeqa
Dividing throughout by the positive number $P(\quad,
\quad)_{\mbox{qm}}^{\mbox{cs}}\/$ and using the expressions of 
eq.~(\ref{coh-coincidence}) we see that the above initial
inequality reduces to the Bell form:
\beqa
-P(\quad,\quad)_{\mbox{qm}}^{\mbox{cs}}\/ \leq\/
P(\theta_1,\theta_2)_{\mbox{qm}}^{\mbox{cs}}
&-&P(\theta_1,\theta_2^{\prime})_{\mbox{qm}}^{\mbox{cs}}
+P(\theta_1^{\prime},\theta_2)_{\mbox{qm}}^{\mbox{cs}}
\nonumber \\
&+&P(\theta_1^{\prime},\theta_2^{\prime})_{\mbox{qm}}^{\mbox{cs}}
-P(\theta^{\prime}_1,\quad)_{\mbox{qm}}^{\mbox{cs}}
-P(\quad,\theta_2)_{\mbox{qm}}^{\mbox{cs}}
 \/\leq\/ 0
\label{chs-ineq-coh}
\eeqa
Thus we see explicitly that coherent states do not violate
the inequalities~(\ref{chs-ineq}). 
The reduction of equation~(\ref{coherent-lemma}) 
to equation~(\ref{chs-ineq-coh}) is possible because coincidence
count rates for coherent states factorize. Further, we emphasize that
it is possible to 
write an inequality corresponding to~(\ref{coherent-lemma}) for any
quantum state but it 
can be reduced to a Bell type form~(\ref{chs-ineq-coh}) only for those 
special states for which coincidence count rates factorize. 
\subsection{The classical states}
We can express any arbitrary state of the 4-mode radiation field
in terms of projections onto coherent states~\cite{klauder}; 
in particular, for a 
state with density matrix $\rho\/$, we have 
\begin{mathletters}
\beqa
&\rho= 
\displayfrac{1}{\pi^4}\int \varphi(z_{1},z_{2},
z_{3},z_{4}) 
\vert z_{1},z_{2},z_{3},z_{4}\rangle
\langle z_{1},z_{2},z_{3},z_{4}\vert 
d^2 z_{1}d^2 z_{2}d^2 z_{3}d^2 z_{4}&
\label{def-diag-coh-4},\\
&\displayfrac{1}{\pi^4}
\int \varphi(z_{1},z_{2},z_{3},z_{4}) 
d^2 z_{1}d^2 z_{2}d^2 z_{3}d^2 z_{4} =1&
\label{norm-diag-coh-4}
\eeqa
\end{mathletters}
The diagonal coherent state distribution function 
$\varphi(z_{1},z_{2},z_{3},z_{4})\/$ describes
the state $\rho\/$.
In quantum optics the  
states of the 4-mode field are classified into
classical and nonclassical types as follows:
A given state is 
classical if the diagonal coherent state distribution function $\varphi\/$
for it 
is nonnegative and nowhere more singular than a delta function. Otherwise the
state is nonclassical.
The function $\varphi\/$ undergoes a point transformation when the state
undergoes a unitary evolution corresponding to a passive
canonical tranformation given by an element of $U(4)$:
\beqa
\varphi(z_{1},z_{2},z_{3},z_{4}) &\rightarrow&
\varphi^{\prime}(z_{1},z_{2},z_{3},z_{4})=
\varphi(z^{\prime}_{1},z^{\prime}_{2},
z^{\prime}_{3},z^{\prime}_{4})
\nonumber \\
\left(\begin{array}{cccc}
z^{\prime}_{1}&z^{\prime}_{2}&z^{\prime}_{3} 
&z^{\prime}_{4}
\end{array} \right)
&=&
\left(
\begin{array}{cccc}
z_{1}&z_{2}&z_{3}&z_{4}
\end{array}\right)
U^{T}\quad, \quad  U \in U(4).
\eeqa
Therefore, the classical or nonclassical nature of a state is 
preserved under such transformations.

Coincidence count rates, and for that matter expectation values of any
observable can be calculated from the diagonal coherent state distribution
function.
For a {\em classical state},
the inequality~(\ref{chs-ineq}) becomes, by multiplication
of all terms in~(\ref{chs-ineq-coh}) by $\varphi\/$ followed
by integration
\beqa
&&\!\!\!\!\!\!
-\int \displayfrac{1}{\pi^{4}}P(\quad,\quad)_{\mbox{qm}}^{\mbox{cs}}
\varphi(z_{1},z_{2},z_{3},z_{4})
d^2 z_{1}d^2 z_{2}d^2 z_{3}d^2 z_{4} \/ \leq\/
\nonumber \\
&&\quad\quad \displayfrac{1}{\pi^{4}}\int 
(P(\theta_1,\theta_2)_{\mbox{qm}}^{\mbox{cs}}
-P(\theta_1,\theta_2^{\prime})_{\mbox{qm}}^{\mbox{cs}}
+P(\theta_1^{\prime},\theta_2)_{\mbox{qm}}^{\mbox{cs}}
+P(\theta_1^{\prime},\theta_2^{\prime})_{\mbox{qm}}^{\mbox{cs}}
-P(\theta^{\prime}_1,\quad)_{\mbox{qm}}^{\mbox{cs}}
\nonumber \\
&&\quad\quad \quad \quad
-P(\quad,\theta_2)_{\mbox{qm}}^{\mbox{cs}})
\varphi(z_{1},z_{2},z_{3},z_{4})
d^2 z_{1}d^2 z_{2}d^2 z_{3}d^2 z_{4}
 \/\leq\/ 0
\eeqa
The above result makes use of the nonnegative nature of 
$\varphi\/$ and will not be true for a nonclassical state.
Using the normalisation~(\ref{norm-diag-coh-4}), and the
nonnegative nonsingular nature of
$\varphi(z_{1},z_{2},z_{3},z_{4})\/$,
 we conclude that a ``classical
state'' will not violate Bell type inequalities defined in
eq.~(\ref{chs-ineq}). Further, since the classical or
nonclassical status of  the 4-mode state is invariant
under passive canonical transformations which form the
group $U(4)\/$, a classical state after undergoing such
transformations will still not violate Bell type
inequalities. On the other hand, the nonclassical states \ie
the states with negative or singular diagonal coherent
state distribution functions can violate these inequalities;
in fact, the violation of such an inequality implies that
the underlying diagonal coherent state distribution function
for the state is negative or singular \ie the state is
nonclassical in the quantum optical sense. 

\subsection{Application to two-mode Gaussian states}
We have seen that typically a minimum of four modes are
required for the analysis of Bell's inequalities; but
if we wish to study two-mode nonclassical states and their
potential to violate Bell-type inequalities then we can
choose the other two modes to be in the vacuum state or in
general in a classical state, then perform a suitable
$U(4)\/$ transformation and proceed with the analysis. If 
we observe any violation, it can then be attributed to the initial
nonclassical two-mode state. This is very
similar to the detection of squeezing, where the mixing of
the squeezed signal with high intensity coherent light is
required in order to measure squeezing.

Consider a 4-mode state with a general centered Gaussian
distribution~\cite{arvind-pramana-95}
\begin{mathletters}
\beqa
&&W(\xi)=\pi^{-4} ({\rm Det} G)^{\frac{1}{2}} 
\exp(-\xi^{T} G \xi), 
\label{4-mode-gauss}
\\
&&\xi^{T}=\left(
\begin{array}{cccccccc}
q_{1}& q_{2}&q_{3}&
q_{4}&p_{1}&p_{2}&
p_{3}&p_{4}
\end{array}\right)
\nonumber\\ 
&&G = \mbox{real symmetric positive definite $8\times 8\/$  matrix}.
\eeqa
\end{mathletters}
Here $q$'s and $p$'s are quadrature components corresponding to 
$a$' and $a^{\dagger}$'s 
($q_{1}=\frac{1}{\sqrt{2}}(a_{1}^{\dagger}+a_{1}),
p_{1}=\frac{i}{\sqrt{2}}(a_{1}^{\dagger}-a_{1})$ etc.).
For $W(\xi)\/$ to represent a quantum mechanical state 
and to be a Gaussian Wigner distribution, $G\/$ has to satisfy 
in addition the condition
\beqa
&G^{-1} + i \beta \geq 0&
\nonumber \\
&\beta = \left( \begin{array}{cc}
 0_{4\times4} & 1_{4\times4}\\
-1_{4\times4} & 0_{4\times4}
\end{array} \right)&
\eeqa
This is  an expression of the uncertainty relations
between the canonically conjugate $q\/$'s and $p\/$'s.
The matrix $V=\frac{1}{2}G^{-1}\/$ is the variance or the noise matrix. 
For a given state, if  the least eigenvalue ${\cal E}_<(V)\/$
of this matrix is less 
than $\frac{1}{2}\/$ then the state is squeezed and 
hence nonclassical~\cite{simon-pra-94}:
\beq
{\cal E}_{<}(V) < \frac{1}{2} \Leftrightarrow \mbox{Squeezing}
\eeq 
We now proceed to study the possibility of violation of the Bell type
inequality~(\ref{chs-ineq}) for Gaussian states and its possible
correlation with squeezing.
To calculate the coincidence count rates
we need to express the $U(4)\/$ matrix corresponding to the
rotation to the  basis defined by the polarisers, as a matrix of $Sp(8,R)$;
we construct it directly
from~(\ref{pol}) 
\beq
U(\theta_1,\theta_2)=
\left(\begin{array}{cccccccc}
\cos{\theta_1}&-\sin{\theta_1}&0&0& 0&0&0&0\\
\sin{\theta_1}&\cos{\theta_1}  &0&0& 0&0&0&0\\
0&0&\cos{\theta_2}&-\sin{\theta_2}& 0&0&0&0\\
0&0&\sin{\theta_2}&\cos{\theta_2}  & 0&0&0&0\\
0&0&0&0 &\cos{\theta_1}&-\sin{\theta_1}&0&0\\
0&0&0&0 &\sin{\theta_1}&\cos{\theta_1}&  0&0\\
0&0&0&0 &0&0&\cos{\theta_2}&-\sin{\theta_2}\\
0&0&0&0 &0&0&\sin{\theta_2}&\cos{\theta_2} 
\end{array}
   \right) 
\eeq
In order to calculate the coincidence count rates, we first
compute the overlaps of the 4-mode Gaussian state with
appropriate vacuum states. The Wigner function for a
single-mode vacuum state is given by
\beq
W_0(q,p)=\displayfrac{1}{\pi}\exp{(-q^2-p^2)}
\eeq
The relevant overlaps can then be written as\footnote[1]{$e_{ij}\/$ is
an $ 8\times8\/$ matrix with $(e_{ij})_{ij} =1\/$ and all other
elements zero; $I_{8 \times 8}\/$ is eight dimensional
identity matrix}
\beqa
\mbox{Tr}(\rho \vert 0 \rangle_{\theta_1} \,{}_{\theta_1}\langle 0 \vert) 
&=& 2 \pi \int W\left(U(\theta_1,0)\, \xi\right)
\/\/W_0(q_{1},p_{1}) \/d \xi \nonumber \\
&=&  2 \sqrt{\mbox{Det}(U^T(\theta_1,0) G U(\theta_1,0) +e_{11}+e_{55})^{-1}}
\nonumber \\
\mbox{Tr}(\rho \vert 0 \rangle_{\theta_2} \,{}_{\theta_2}\langle 0 \vert) 
&=& 2 \pi \int W\left(U(0,\theta_2)\, \xi \right)
\/\/W_0(q_{3},p_{3}) \/d \xi \nonumber \\
&=&  2 \sqrt{\mbox{Det}( U^T(0,\theta_2) G U(0,\theta_2)+e_{33}+e_{77})^{-1}}
\nonumber \\
\mbox{Tr}(\rho \vert 0 \rangle_{\theta_1}
\,{}_{\theta_1}\langle 0 \vert \vert 0 \rangle_{\theta_2}
\,{}_{\theta_2}\langle 0 \vert) 
&=& (2 \pi)^2 \int W\left(U(\theta_1,\theta_2)\, \xi\right)
\/\/W_0(q_{1},p_{1})W_0(q_{3},p_{3})
\/d \xi \nonumber \\
&=& 4 \sqrt{\mbox{Det} (U^T(\theta_1,\theta_2) G
U(\theta_1,\theta_2)+e_{11}+ e_{33}+e_{55}+e_{77})^{-1}}
\nonumber \\
\mbox{Tr}(\rho \vert 0 0 \rangle_{{\bf k}} \,{}_{\bf k}\langle 0 0\vert) 
&=& (2 \pi)^2 \int W\left(\xi \right)
\/\/W_0(q_{1},p_{1})\/\/W_0(q_{2},p_{2}) \/d \xi \nonumber \\
&=& 4 \sqrt{\mbox{Det}( G+e_{11}+e_{22}+e_{55}+e_{66})^{-1}}
\nonumber \\
\mbox{Tr}(\rho \vert 0 0 
\rangle_{{\bf k}^{\prime}} \,{}_{{\bf k}^{\prime}}\langle 0 0\vert) 
&=& (2 \pi)^2 \int W\left(\xi \right)
\/\/W_0(q_{3},p_{3})\/
W_0(q_{4},p_{4}) \/d \xi \nonumber \\
&=&   4 \sqrt{\mbox{Det}( G+e_{33}+e_{44}+e_{77}+e_{88})^{-1}}
\nonumber \\
\mbox{Tr}(\rho \vert 0 0 \rangle_{{\bf k}} \,{}_{{\bf k}}\langle 0 0 \vert
\vert 0 0 \rangle_{{\bf k}^{\prime}}
 \,{}_{{\bf k}^{\prime}}\langle 0 0 \vert) 
&=&(2 \pi)^4 \int W\left(\xi\right)
\/\/W_0(q_{1},p_{1})W_0(q_{2},p_{2})W_0(q_{3},p_{3})
W_0(q_{4},p_{4})\/ d \xi 
\nonumber \\
&=& 16 \sqrt{\mbox{Det}(G+I_{8\times 8})^{-1}}
\nonumber \\
\mbox{Tr}(\rho \vert 0 \rangle_{\theta_1}
\,{}_{\theta_1}\langle 0 \vert 
\vert 0 0 \rangle_{{\bf k}^{\prime}} \,{}_{{\bf k}^{\prime}}\langle 0 0\vert) 
&=& (2 \pi)^3 \int W\left(U(\theta_1,0)\, \xi \right)
\/\/W_0(q_{1},p_{1}) 
W_0(q_{3},p_{3})W_0(q_{4},p_{4})
\/d \xi \nonumber \\
&=&  8 \sqrt{\mbox{Det}( U^T(\theta_1,0) G U(\theta_1,0)+
I_{8\times 8}-e_{22}-e_{66})^{-1}} 
\nonumber \\
\mbox{Tr}(\rho \vert 0 \rangle_{\theta_2}
\,{}_{\theta_2}\langle 0 \vert 
\vert 0 0 \rangle_{\bf k} \,{}_{\bf k}\langle 0 0\vert) 
&=& (2 \pi)^3 \int W\left(U(0,\theta_2)\, \xi \right)
\/\/W_0(q_{3},p_{3}) 
W_0(q_{1},p_{1})W_0(q_{2},p_{2})
\/d \xi \nonumber \\
&=&  8 \sqrt{\mbox{Det}( U^T(0,\theta_2) G U(0,\theta_2)+
I_{8\times 8}-e_{44}-e_{88})^{-1}} \nonumber \\
\label{gauss-overlap}
\eeqa
Using these overlap integrals, 
and the definitions~(\ref{A-operators})
we can immediately compute the quantum
mechanical coincidence count rates 
\beqa
P(\theta_1,\theta_2)_{\rm qm}^{\mbox{gauss}} &=& 
1 -\mbox{Tr}(\rho \vert 0 \rangle_{\theta_1}
\,{}_{\theta_1}\langle 0 \vert)-\mbox{Tr}(\rho \vert 0 \rangle_{\theta_2}
\,{}_{\theta_2}\langle 0 \vert)  
+\mbox{Tr}(\rho \vert 0 \rangle_{\theta_1}
\,{}_{\theta_1}\langle 0 \vert \vert 0 \rangle_{\theta_2}
\,{}_{\theta_2}\langle 0 \vert)
\nonumber \\
P(\theta_1,\quad)_{\rm qm}^{\mbox{gauss}} &=&
1 -\mbox{Tr}(\rho \vert 0 \rangle_{\theta_1}
\,{}_{\theta_1}\langle 0 \vert)-
\mbox{Tr}(\rho \vert 00 \rangle_{{\bf k} ^{\prime} }
\,{}_{{\bf k}^{\prime}}\langle 00 \vert)  
+\mbox{Tr}(\rho \vert 0 \rangle_{\theta_1}
\,{}_{\theta_1}\langle 0 \vert \vert 0 0 \rangle_{{\bf k}^{\prime}}
\,{}_{{\bf k}^{\prime}}\langle 0 0 \vert)
\nonumber \\
P(\quad,\theta_2)_{\rm qm}^{\mbox{gauss}} &=&
1 -\mbox{Tr}(\rho \vert 0 0\rangle_{\bf k}
\,{}_{\bf k}\langle 0 0 \vert)-
\mbox{Tr}(\rho \vert 0 \rangle_{\theta_2}
\,{}_{\theta_2}\langle 0 \vert)  
+\mbox{Tr}(\rho \vert 0 0\rangle_{{\bf k}}
\,{}_{{\bf k}}\langle 0 0 \vert \vert 0 \rangle_{\theta_2}
\,{}_{\theta_2}\langle 0 \vert)
\nonumber \\
P(\quad,\quad)_{\rm qm}^{\mbox{gauss}} &=&
1 -\mbox{Tr}(\rho \vert 0 0 \rangle_{{\bf k}}
\,{}_{{\bf k}}\langle 0 0\vert)-\mbox{Tr}(\rho \vert 0 0 \rangle_{{\bf k}^{\prime}}
\,{}_{{\bf k}^{\prime}}\langle 0 0 \vert)  
+\mbox{Tr}(\rho \vert 0 0\rangle_{{\bf k}}
\,{}_{{\bf k}}\langle 0 0 \vert \vert 0 0 \rangle_{{\bf k}^{\prime}}
\,{}_{{\bf k}^{\prime}}\langle 0 0 \vert)
\label{gauss-coincidence}
\nonumber \\
\eeqa
The equations~(\ref{gauss-overlap}), can now be used to analyse any
Gaussian Wigner state to see if it violates the inequality~(\ref{chs-ineq}).
Having set up the general formalism for the family of centered Gaussian Wigner 
states, we now consider examples of $G\/$ which lead to violation of  
the inequality~(\ref{chs-ineq}). Consider 
\beqa
G &=& U^{-1} S^T G_{0} S U
\nonumber \\
G_{0}&=& \kappa \/ I_{8\times 8},\quad 0 \leq \kappa \leq 1.
\nonumber \\
\kappa&=& \tanh{\displayfrac{\beta}{2}},\quad \beta =
\displayfrac{\hbar \omega}{k T}
\eeqa
Here $\kappa=1\/$ implies zero temperature and $ \kappa < 1\/$
corresponds to some finite temperature, $S\/$ is a
4-mode squeezing symplectic transformation, which is a
$Sp(8,\Re)\/$ matrix, and $U\/$ is a passive symplectic
$U(4)\/$ transformation whose role is to produce 
entanglement. As an example, we start with a
state in which the modes $a_{1}\/$ and $a_{4}\/$
are squeezed by equal and opposite amounts $u\/$ and the
modes $a_{2}\/$ and $a_{3}\/$ are squeezed by 
equal and opposite amounts $v\/$, and the entanglement is
``maximum''. This corresponds to the
choices\footnote[1]{This $S\/$ matrix is constructed from two
two-mode squeezing transformations studied in detail 
in~\cite{arvind-pra-95}, 
the first being the one which squeezes modes 1 and
4 by equal and opposite amounts $u\/$, whereas the second
one squeezes the modes 2 and 3 by equal and opposite
amounts $v\/$}
\beqa
S=\left(\begin{array}{cccccccc}
e^{-u}&0&0&0&0&0&0&0\\
0&e^{v} &0&0&0&0&0&0\\
0&0&e^{-v}&0&0&0&0&0\\
0&0&0&e^{u} &0&0&0&0\\
0&0&0&0&e^{u} &0&0&0\\
0&0&0&0&0&e^{-v}&0&0\\
0&0&0&0&0&0&e^{v} &0\\
0&0&0&0&0&0&0&e^{-u}
\end{array}
   \right),\/ 
U=\displayfrac{1}{2}\left(\begin{array}{rrrrrrrr}
  1&1&1&1&0&0&0&0\\
-1&1&-1&1&0&0&0&0\\
-1&-1&1&1&0&0&0&0\\
1&-1&-1&1&0&0&0&0\\
  0&0&0&0&1&1&1&1\\
0&0&0&0&-1&1&-1&1\\
0&0&0&0&-1&-1&1&1\\
0&0&0&0&1&-1&-1&1      
\end{array}
   \right)
\nonumber \\
\mbox{}
\eeqa
For this particular class of centered Gaussian Wigner states the function 
(dependences on $u\/$ and $v\/$ are left implicit)
\beqa
f(\theta_1,\theta_2, \theta_1^{\prime}, \theta_2^{\prime})
=
P(\theta_1,\theta_2)_{\rm qm}^{\mbox{gauss}}-
P(\theta_1,\theta^{\prime}_2)_{\rm qm}^{\mbox{gauss}}+
P(\theta^{\prime}_1,\theta_2)_{\rm qm}^{\mbox{gauss}}+
\nonumber \\
P(\theta^{\prime}_1,\theta^{\prime}_2)_{\rm qm}^{\mbox{gauss}}
-P(\theta^{\prime}_1,\quad)_{\rm qm}^{\mbox{gauss}}-
P(\quad,\theta_2)_{\rm qm}^{\mbox{gauss}}
\eeqa
can be calculated. 
\begin{figure}
\epsfxsize=16.5cm 
\epsfbox{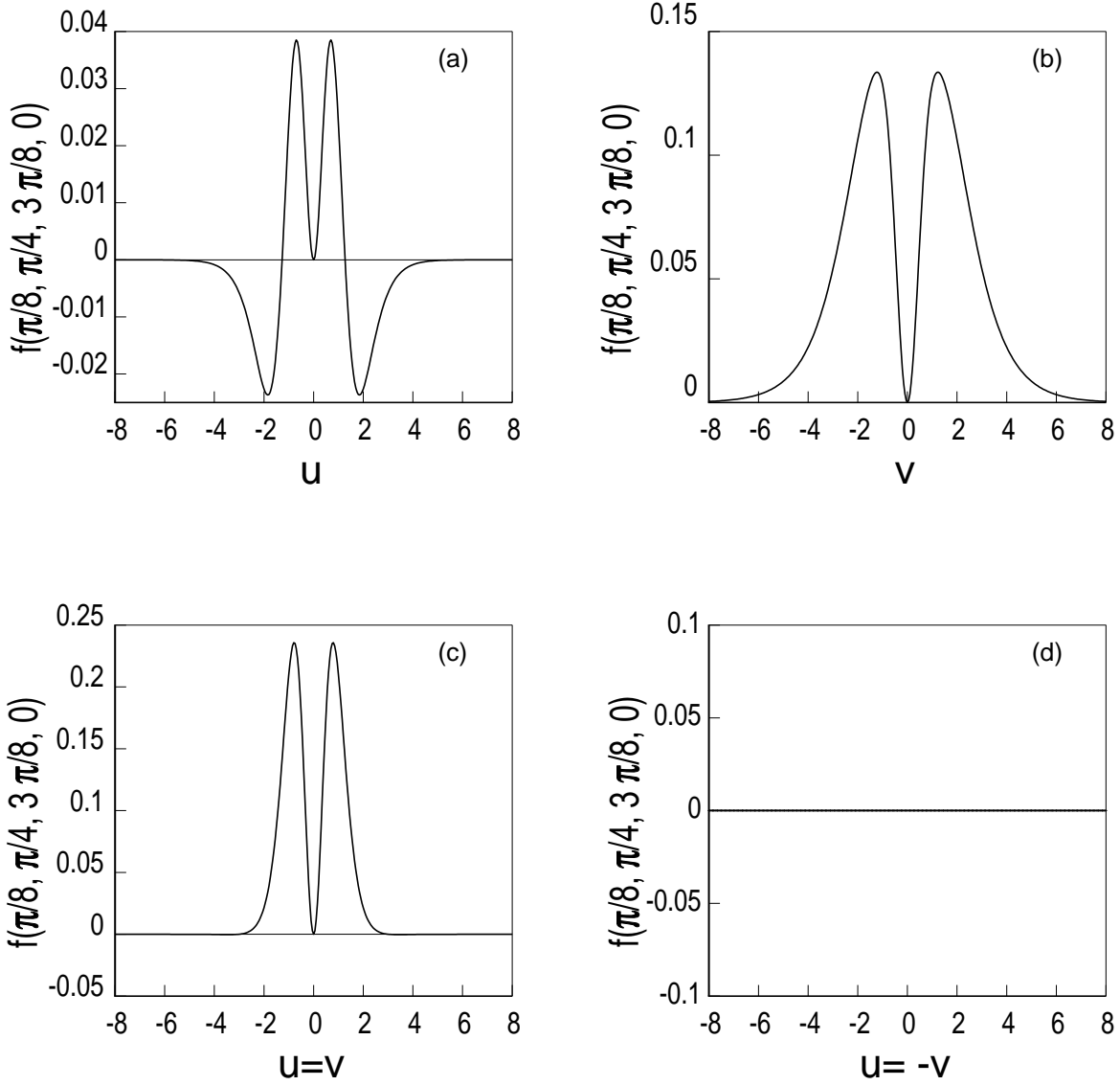}
\caption{Violation of Bell type inequality for
states with centered Gaussian Wigner distributions
representing a 4-mode squeezed vacuum. 
(a) $v=0\/$ \ie Modes 1 and 4 are squeezed by equal
and opposite amounts $u\/$ 
(b) $u=0\/$ \ie Modes 2 and 3 are squeezed
by equal and opposite amounts
(c) $u=v\/$ \ie all modes are squeezed.
(d) $u=-v\/$ \ie all modes are squeezed. }
\end{figure}
\begin{figure}
\epsfxsize=15.5cm 
\epsfbox{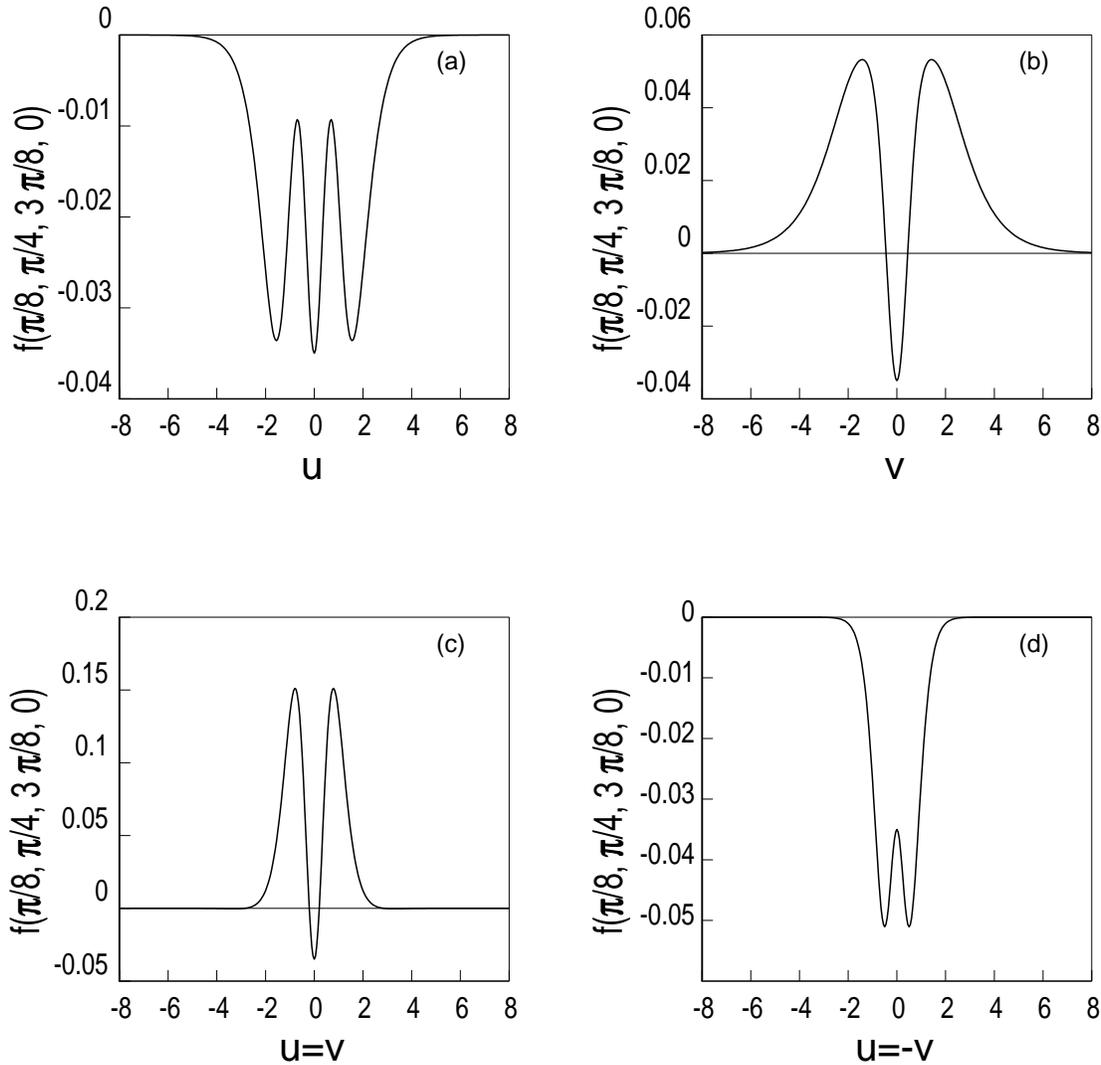}
\caption{Violation of Bell type inequality for
states with centered Gaussian Wigner distributions
representing 4-mode squeezed thermal states with 
the same temperature for all the modes. The choice of
parameters for figures (a), (b), (c) and (d) is the same as that
for Figure~2.}
\end{figure}
Though one ought to search over all values of the angles
$\theta_1, \theta_2, \theta_1^{\prime}\/$ and $\theta^{\prime}_2\/$
to look for possible violations of the 
inequality~(\ref{chs-ineq}), motivated by the choice of
angles for the two-photon state, we restrict ourselves to
that same choice, and plot the function
$f(\displayfrac{\pi}{8},\displayfrac{\pi}{4} ,3
\displayfrac{\pi}{8},0)\/$
for various combinations of the squeeze parameters in
Figures~2 and~3.  In Figure~2 we have considered the case
$\kappa=1\/$ \ie the state under consideration is the squeezed
vacuum; whereas for Figure~3 we have
$\kappa=0.75\/$
which corresponds to a squeezed thermal state.  For the case of
the squeezed
vacuum, a clear violation is demonstrated in figures~2(b)
and~2(c), for a considerable range of squeeze parameters. No
violation is seen for the parameters chosen in
Figure~2(d), and Figure~2(a) shows very little violation. 
However, for these parameter values for
squeezing the violation might occur for some other values
of angles $\theta_1, \theta_2,
\theta^{\prime}_1\/$ and $\theta_2^{\prime}\/$.  
Figure~3 has similar features as Figure~2 but we see as
expected that the amount of violation has diminished when the
initial state is thermal instead of vacuum; as a consequence the 
small amount of violation which was present in Figure~2(a) has disappeared in 
Figure~3(a). As a further study,  
in the context of Gaussian Wigner states, it will be 
interesting to see the effect of phase space displacement on the 
violation of these inequalities; this  will be taken up elsewhere.
\section{Concluding remarks}
We have developed the machinery for analysing the violation
of Bell type inequalities for a general state of
the 4-mode radiation field in a setup of the type described in
Fig~1.  A classical state in the quantum optical sense 
always obeys these inequalities
while a nonclassical state may violate them, possibly after
a $U(4)\/$ transformation. Starting with
a general nonclassical state, we subject it to a general
unitary evolution corresponding to passive canonical
transformations $U(4)\/$ before we look for the violation of
Bell type inequalities. It may turn out that a given
nonclassical state does not violate Bell type inequalities
but some $U(4)\/$ variant of it does. 

A pure quantum mechanical state of a composite system 
is said
to be entangled if we are not able to express it as a
product of two factors, one belonging to each subsystem.
Such states have nontrivial quantum correlations and can
lead to the violation of suitable Bell's inequalities.
The simplest quantum optical system for which the notion of entanglement 
can be introduced is the two-mode field.
The 
group of passive canonical transformations in this case  
is $U(2)\/$; its elements, though incapable of  producing  or
destroying  nonclassicality are capable of
entangling(disentangling) originally unentangled(entangled)
states.
  As an  simple example we
choose the nonclassical state $\vert 1 \rangle \vert 1
\rangle \/$ which
is not entangled; by a simple $U(2)\/$
transformation ${\displaystyle e}^{\displaystyle
\displayfrac{i\pi}{4}
(a_1^{\dagger}a_2+ a_2^{\dagger}a_1)}\,$ it becomes
the  entangled state $\displayfrac{1}{\sqrt{2}}(\vert 2
\rangle
\vert 0 \rangle + \vert 0 \rangle \vert 2 \rangle)\/$.
However, coherent states are not entanglable in this
way!  If we start with a two-mode coherent state $\vert
z_1, z_2 \rangle\/$, it is clearly not entangled
\ie the state can be written as a product with one factor 
belonging to one mode and the other to the other mode.
Under a $U(2)\/$ transformation this property is
maintained.  Classical states are statistical mixtures of
coherent states and under $U(2)\/$ transform again to such
mixtures of coherent states. Such a mixture can definitely
have correlations which are purely classical, but it cannot
have truly quantum mechanical 
entanglement.  Thus classical states are  to be regarded as 
not entangled,
and they remain so under passive $U(2)\/$ transformations.
However this is in general not true for a nonclassical
nonentangled state which may get entangled under a suitable
$U(2)\/$ transformation. It is a straightforward matter to
generalise the above statements to $n\/$ mode systems where the
group of passive canonical transformations is $U(n)\/$.
There are several ways to quantify entanglement; for pure
states it is unambiguous: if the reduced density
matrices for the subsystems involved are also pure states
then the original state is not entangled.  On the other hand, if in
the process of partial tracing, some information is lost then the
original state is entangled. The generalisation to mixed
states is nontrivial  but is conceptually simple; we
have to separate classical correlations from the quantum
mechanical ones and this may not always be easy to do.
However, as we saw for the case of classical states, we can
sometimes easily conclude that a given state is nonentangled.

The above conclusions have an interesting bearing  on
the work on violation of Bell-type inequalities with beams 
originating from independent
sources~\cite{yurke-prl-92}~\cite{yurke-pra-92}~\cite{zukowski-prl-92}.
These experiments  
take two beams from two independent sources, pass them 
through some passive optical elements and show that the 
Bell-type inequalities are violated. The first conclusion we can 
draw from 
our analysis is that it must be the  quantum optical nonclassicality of 
one of the beams in this experiment 
which has been converted into entanglement by the $U(4)\/$ transformation
and hence led to the violation. Secondly,
if the original beams were
quantum optically classical, no matter what one does, no violation will
be seen. 

In our analysis, we have not distinguished between
strengths of coincidences. The coincidence counter
registers a count when simultaneously each detector detects
one or more photons. This is the reason why we chose the
operators $A$'s to have eigen values $0\/$ and $1\/$.  In
this sense, the measurements involved here are not refined.
It would be interesting to further generalise the analysis
by considering somewhat refined measurements where to some
extent coincidences are
distinguished on the basis of their strengths. However, the
relevant operators in this context may be unbounded; and it
is well known that the formulation of Bell type
inequalities for such operators, though desirable, is
nontrivial.

We have compared quantum optical nonclassicality with
violation of Bell's inequalities. When a state is
nonclassical in the quantum optical sense, it does not
allow a classical description based on an ensemble of
solutions of Maxwell's equations, which is a very specific
classical theory. On the other hand violation by a state of
a Bell type inequality rules out any possibility of
describing it by any general local ``classical'' hidden
variable theories. Therefore, it is understandable that
quantum optical nonclassicality is a necessary but not a
sufficient condition for the violation of Bell's
inequalities. This disparity is partially compensated for
by the freedom to perform passive canonical transformations
on a nonclassical state before looking for violation of
Bell's inequalities though it is not obvious whether this
freedom completely removes this discrepancy. On the other
hand, if a state
obeys Bell's inequalities, it may still not allow a 
``classical'' 
description. Therefore, we need a complete set of Bell's
inequalities capturing the full content of the locality
assumption. These and related aspects will be explored 
elsewhere. 

\end{document}